\documentclass[aps, twocolumn,superscriptaddress,longbibliography]{revtex4-2}
\usepackage{epsfig,amssymb,amsmath,amsthm,amsfonts,amsbsy,mathrsfs}
\usepackage{graphicx}
\usepackage{color}
\usepackage{titlesec}
\usepackage{tikz}
\usetikzlibrary{arrows,shapes,chains}

\usepackage{pifont}

\definecolor{cream}{RGB}{222,217,201}

\begin{document}
\title{Migration of active particle in mixtures of rigid and flexible rings}
\author{Meng-Yuan Li}
\affiliation{Key Laboratory of Advanced Optoelectronic Quantum Architecture and Measurement (MOE), School of Physics, Beijing Institute of Technology, Beijing, 100081, China}
\author{Ning Zheng}
\email{ningzheng@bit.edu.cn}
\affiliation{Key Laboratory of Advanced Optoelectronic Quantum Architecture and Measurement (MOE), School of Physics, Beijing Institute of Technology, Beijing, 100081, China}
\author{Yan-Wei Li}
\email{yanweili@bit.edu.cn}
\affiliation{Key Laboratory of Advanced Optoelectronic Quantum Architecture and Measurement (MOE), School of Physics, Beijing Institute of Technology, Beijing, 100081, China}
\date{\today}

\begin{abstract}
The migration of active particles in slowly moving, crowded, and heterogeneous media is fundamental to various biological processes and technological applications, such as cargo transport. In this study, we numerically investigate the dynamics of a single active particle in a medium composed of mixtures of rigid and flexible rings. We observe a non-monotonic dependence of diffusivity on the relative fraction of rigid to flexible rings, leading to the identification of an optimal composition for enhanced diffusion. This long-time non-monotonic diffusion, likely resulting from the different responses of the active particle to rigid and flexible rings, is coupled with transient short-time trapping. The probability distribution of trapping durations is well described by the extended entropic trap model. We further establish a universal relationship between particle activity and the optimal rigid-to-flexible ring ratio for diffusion, which aligns closely with our numerical results.
\end{abstract}
\maketitle

Active matter can self-propel and harvest energy from its environment, driving itself out of equilibrium~\cite{Watkins2012,Marchetti2013,Hartmann2019,Hauser2015,Ramaswamy2010, WuCSR}. This unique property plays a critical role in various processes, including bioremediation~\cite{Sarah2013,Gao2014}, microbial drug delivery~\cite{Pelin2019,RN85}, gene transcription~\cite{Cramer}, and environmental remediation~\cite{Joanna2018}. For instance, pathogens migrate through tissues during infections~\cite{HarmanPNAS,Kelly2013}, and engineered bacteria can transport therapeutic agents into tumors, offering innovative strategies for drug delivery~\cite{drugdeliver,drugdelivery}. Unlike passive systems, active cargo carriers overcome numerous limitations, enabling new functionalities~\cite{nl2008,Park2021}. Understanding the physical mechanisms governing active matter in complex environments is therefore essential for advancing biological insights and novel nanotechnological applications.

It has been established that biological or synthetic active matter often exhibits anomalous sub- or super-diffusive behavior due to the interplay between its persistent motion and environmental constraints~\cite{C9CP04498A,Praveen2023,Theresa2024,RN61,Ehsan2022}. For instance, in soft, narrow channels such as blood vessels or microfluidic systems~\cite{LJH2022}, the clustering behavior of active particles depends on channel size and particle activity~\cite{KnippenbergPRL}. Such confinements also result in heterogeneous transport, characterized by transitions between localized trapping and bursts of rapid motion, leading to non-Gaussian diffusion profiles~\cite{Wu2017,Ann2024}. In porous media, diffusion is influenced by surface scattering~\cite{RN29,Theresa2018,Theresa2024}, obstacle spatial distribution~\cite{Fergus2023,Giorgio2017}, and the shape, size, and stiffness of the active particles~\cite{ActivePolymer}. For example, bacteria encountering obstacles of comparable size exhibit enhanced diffusion through surface scattering at low densities but reduced diffusivity due to frontal collisions at high densities~\cite{bacteria2019}. Navigation mechanisms, such as running and rolling, enable microorganisms to adapt to environmental changes, improving diffusion efficiency and facilitating behaviors like foraging or avoiding harmful substances~\cite{Anupam2017,Watkins2012}. In fixed porous media, bacteria exhibit intermittent trapping and hopping, with optimal spreading determined by the relationship between their run length and pore size, or by their tumbling rate and velocity~\cite{Tapomoy2019, Christina2021,Lazaro2021,Ehsan2022}. Additionally, the nature of obstacle distributions—whether orderly or disordered—along with particle activity properties, such as run length and chirality, can jointly affect transport, leading to optimal diffusion conditions~\cite{RN30,Leila2020,Chung2024,RN27}. Similar non-monotonic diffusion has been observed in polymer solutions, where the interplay between particle persistence and solution viscosity plays a critical role~\cite{Yunfei2019}.

Previous studies predominantly focused on the migration of active matter in systems with homogeneous and fixed obstacles. However, in real systems, the media can be much more complex. For instance, $Listeria$ $monocytogenes$ can penetrate the cell membrane, enter the cytoplasm, and interact with moving actin and cytoskeletal components to facilitate intracellular movement~\cite{David2015,Pascale2011,Keith2007,Listeriaspread}. In soil, bacteria navigate a heterogeneous matrix composed of sand, silt, clay, organic matter, plant roots, fungal hyphae, and microorganisms. Soil pores serve as both habitats and pathways, but bacterial motility is hindered by barrier particles and binding organic matter, necessitating adaptive strategies for nutrient acquisition and ecological interactions~\cite{YANG2018112,Aroney2021,Li2024}. Fundamental questions about the intrinsic link between activity and the optimal spreading of active particles in complex environments remain elusive.

In this work, we numerically investigate the dynamics of a single active Brownian particle diffusing through slowly moving, crowded mixtures of rigid and flexible rings. Our results reveal a non-monotonic dependence of the particle's diffusivity-and hence the existence of optimal diffusion-on the relative fraction of rigid versus flexible rings across a broad range of persistence times and active particle velocities. This non-monotonic behavior in long-time diffusivity is closely linked to the transient short-time trapping of the active particle by neighboring rings. We theoretically extend the entropic trap model to rationalize the probability distribution of trapping durations and demonstrate that the combined effects of activity and the fraction of rigid versus flexible rings account for the observed non-monotonic behavior. Furthermore, we propose a simple yet universal relationship for predicting optimal diffusion conditions based solely on the particle's persistence time and active velocity, which aligns well with our numerical results.

\section*{Model an Methods}

As illustrated in Fig.~\ref{fig:msd} (a) inset, we study the dynamics of a single self-propelled particle diffusing in a crowded environment composed of a mixture of rigid and flexible rings, in two dimensions. The rigid rings are modeled as rigid bodies, while the flexible ones are bead-spring chains, each consisting of $N_{\rm bead}=100$ beads. The non-bonded interactions between beads (denoted by B) and between beads and active particle (denoted by A) are described by the purely repulsive Weeks–Chandler-Andersen (WCA) potentials \cite{WCA}, given by $U_{\mathrm{nb}}(r_{\mathrm{nb}})=4 \epsilon\left[(\sigma_{\alpha \beta} / r_{\mathrm{nb}})^{12}-(\sigma_{\alpha \beta} / r_{\mathrm{nb}})^6+c\right]$ for $r_{\mathrm{nb}}<r_{\text {cut }}=2^{1 / 6} \sigma_{\alpha \beta}$ and $U_{\mathrm{nb}}(r_{\mathrm{nb}})=0$ otherwise. Here, $r_{\rm nb}$ represents the separation between two beads or between a bead and the active particle. The constant $c$ ensures ${{U}_{\text{nb}}}(r_{\rm cut})=0$. $\alpha ,\beta \in \left\{A,B\right\}$, with ${{{\sigma }_{\mathrm{AB}}}}/{{{\sigma }_{BB}}}=1.5$ and $ {{{\sigma }_{\mathrm{AA}}}}/{{{\sigma }_{\mathrm{BB}}}}=2$. ${{\sigma }_{\mathrm{BB}}}$, $\epsilon $ and ${{\tau }_{c}}=\sqrt{{m\sigma _{\mathrm{BB}}^{2}}/{\epsilon }\;}$ are chosen as the units of length, energy and time, respectively. 
The bond interaction between nearby beads with separation $r_{\rm b}$ for flexible rings is modeled by the finite extensible nonlinear elastic (FENE) potential \cite{FENE}, $U_{\mathrm{FENE}}(r_{\mathrm{b}})=-\frac{1}{2} k_b R_0{ }^2 \ln \left[1-(r_{\mathrm{b}} / R_0)^2\right]$, where $k_b=30 \varepsilon / \sigma_{\mathrm{BB}}^2$ and $R_0=1.5 \sigma_{\mathrm{BB}}$.

The dynamics of beads of flexible rings is described by the Langevin equation:
$m \frac{\mathrm{d}^2 \mathbf{r}_i}{\mathrm{d} t^2} = -\gamma \frac{\mathrm{d} \mathbf{r}_i}{\mathrm{d} t} - \nabla U_i^{\prime} + \sqrt{2 k_{\mathrm{B}} T \gamma} \xi(t)$. $U_i^{\prime}$ denotes the total interaction of bead $i$ with other beads in flexible rings. The dynamics of rigid rings follow the same Langevin equation, with the interaction arising solely from non-bonded interactions between different rings, and the position being the center of mass of the rigid rings. $\xi$ is a Gaussian white-noise, satisfying $\left\langle \xi \left( t \right) \right\rangle =0,\left\langle \xi \left( t \right)\cdot \xi \left( t\text{ }\!\!'\!\!\text{ } \right) \right\rangle =\mathbf{I}\delta \left( t-t\text{ }\!\!'\!\!\text{ } \right)$ with $\mathbf{I}$ the unit tensor. The damping coefficient ${\gamma }=0.01$. The temperature is maintained at $T = 0.1$ using a Nosé-Hoover thermostat. This low temperature induces sluggish dynamics in both the rigid and flexible rings, thereby emphasizing the influence of active particles.

The dynamics of an active particle is given by the overdamped equation of motion: $\dot{\mathbf{r}} = -\frac{\nabla U_{\text{nb}}}{\gamma} + {v}_a \cdot \mathbf{n} + \sqrt{2 \gamma k_B T}\xi(t)$, where the polarity vector $\mathbf{n}=\left( \cos \theta ,\sin \theta  \right)$, and $\dot{\theta} = \sqrt{2 D_r}\eta(t)$.  $D_r$ is the rotational diffusion coefficient resulting in the persistence time $\tau_p=1/D_r$. Larger values of $\tau_p$ correspond to more persistent dynamics. $\tau_p$, along with the active velocity ${{v}_{a}}$, controls the persistent length or the Péclet number $Pe=v_a\tau_p$. $\eta$ is again a Gaussian white-noise with zero mean and unit variance, i.e., $\left\langle \eta \left( t \right) \right\rangle =0,\left\langle \eta \left( t \right)\cdot \eta \left( t\text{ }\!\!'\!\!\text{ } \right) \right\rangle =\delta \left( t-t\text{ }\!\!'\!\!\text{ } \right)$.

 Extensive simulations were performed under periodic boundary conditions with a system containing a total of 100 non-overlapping rigid and flexible rings. The effective packing fraction of the rings is $\phi= 0.87\pm 0.025$, calculated using the average gyration radius of rings $R_g=\left\langle\sqrt{\frac{1}{N_{\text {bead }}} \sum_i\left(R_i-R_c\right)^2}\right\rangle$~\cite{Rg}, where $R_c$ is the center of mass of the $i$-th ring and $\langle...\rangle$ denotes the average over different rings and different configurations. All results were averaged over at least 10 independent runs. We varied the relative fraction of rigid rings $f_{\rm {rigid}}$, the persistence time, and the active velocity to investigate the influence of activity and environment on the diffusion of the active particles.
 \section*{Results}
 \subsection*{Non-monotonic dynamics}
\begin{figure}[!t]
 \centering
 \includegraphics[angle=0,width=0.48\textwidth]{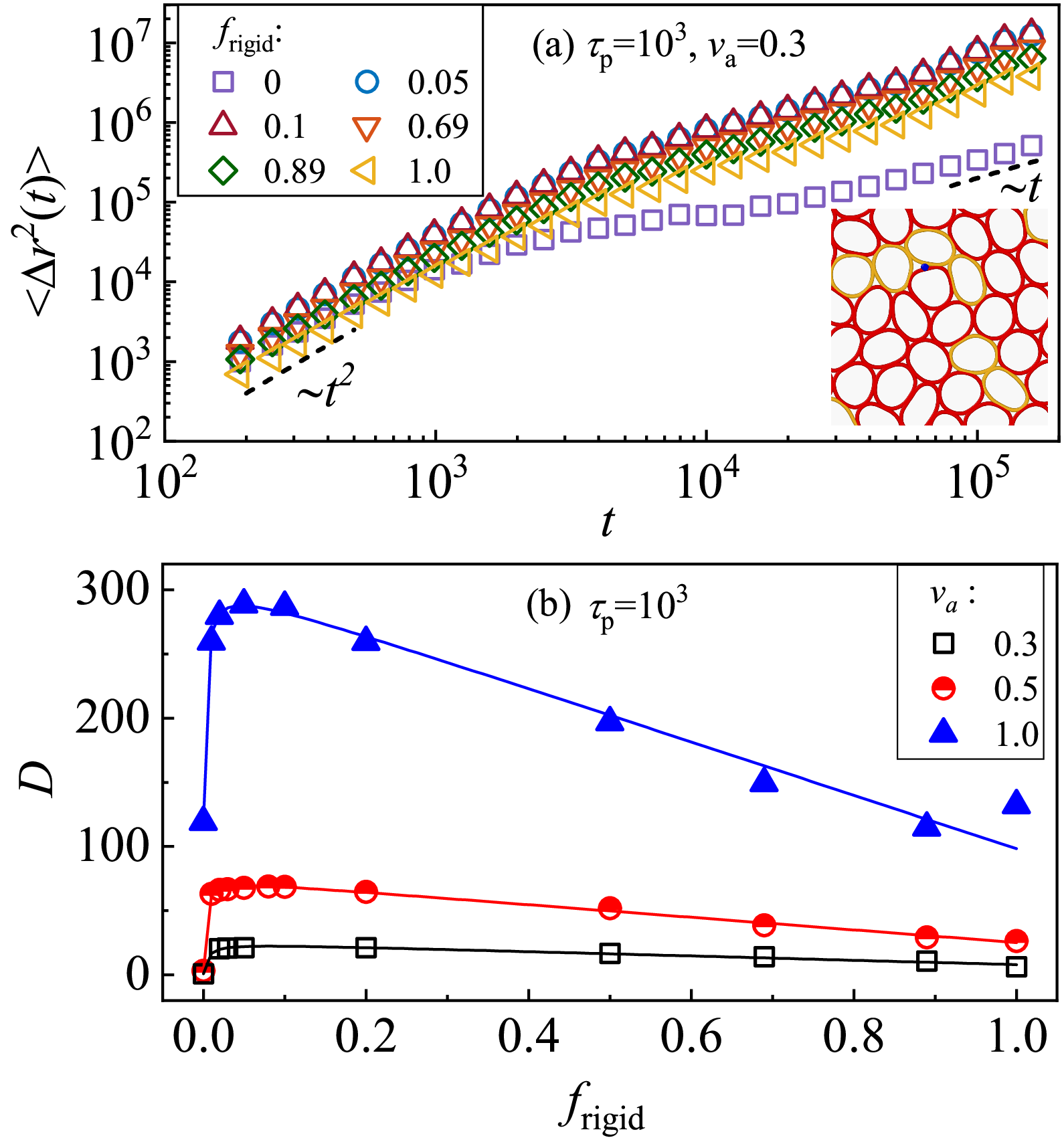}
 \caption{(a) Time dependence of the mean square displacement for a single active particle with $\tau_p = 10^3$ and $v_a = 0.3$, shown for various fractions $f_{\rm{rigid}}$ of rigid rings. The inset presents a snapshot of a system with $f_{\rm{rigid}} = 0.2$. In the inset, rigid rings are colored yellow, flexible rings red, and the small blue disc represents the active particle. (b) Diffusion coefficient as a function of $f_{\rm{rigid}}$ for different values of $v_a = 0.3, 0.5, 1.0$. The lines in (b) are fits by $(a_1 f_{\text {rigid }}^2+a_2 f_{\text {rigid }}+a_3) /(f_{\text {rigid }}+a_4)$ fitting parameters.
}
\label{fig:msd}

\end{figure}

The motion of a single active particle in a mixture of rigid and flexible rings is characterized by its mean square displacement (MSD), $\langle\Delta\mathbf{r}^2(t)\rangle = \langle|\mathbf{r}(t) - \mathbf{r}(0)|^2\rangle$, where $\langle\ldots\rangle$ denotes the ensemble average. Figure~\ref{fig:msd}(a) illustrates the dependence of MSD on the fraction of rigid rings for $\tau_p=10^3$ and $v_a=0.3$. At short times, we observe ballistic behavior, characterized by $\langle\Delta\mathbf{r}^2(t)\rangle\sim t^2$, with the duration of the ballistic motion corresponding to the $\tau_p$ timescale, as expected. At longer times, the motion of the active particle becomes diffusive, allowing for the definition of the diffusion coefficient $D = \frac{1}{4}\mathop {\lim }\limits_{t \to \infty } \frac{d\langle \Delta \mathbf{r}^2(t)\rangle}{dt}$~\cite{Chung2024,Bi_prx}. Moreover, the dynamics of the active particle vary with the fraction of rigid rings. For instance, at $f_{\rm rigid}=0$, sub-diffusive behavior is observed at intermediate times in the MSD, indicating that the active particle is constrained by the flexible rings. These constraining effects are more pronounced when the active velocity $v_a$ is lower, causing the MSD to reach a plateau for an extended period. Intriguingly, introducing a small fraction of rigid rings (e.g., $f_{\rm rigid}=0.05$) accelerates the dynamics and eliminates the constraints, as shown by the comparison of the MSD between $f_{\rm rigid}=0$ and $f_{\rm rigid}=0.05$ in Fig.~\ref{fig:msd}(a). Specifically, the MSD shifts upward, and the sub-diffusive behavior disappears at $f_{\rm rigid}=0.05$. However, as the fraction of rigid rings further increases, the MSD shifts downward, indicating a slowdown in the dynamics. These illustrate the non-monotonic dependence of the active particle's dynamics on the fraction of rigid rings.

The non-monotonic dependence of dynamics on the fraction of rigid rings is more pronounced in Fig.~\ref{fig:msd}(b), where we plot the diffusion coefficient $D$ as a function of $f_{\rm rigid}$ for different values of $v_a$ at a fixed $\tau_p=10^3$. As expected, a higher value of $v_a$ results in a larger $D$. However, regardless of the specific value of $v_a$, $D$ initially increases and then decreases with increasing $f_{\rm rigid}$, peaking around $f_{\rm rigid}=0.05$. This non-monotonic behavior is also observed for $\tau_p=10^2$, as illustrated in supplementary Fig. S1 of the supplementary information (SI)~\cite{SM}, but it disappears for $\tau_p=10$ with $v_a=0.1$ (supplementary Fig. S3~\cite{SM}), suggesting that it arises from persistent motion. 

\begin{figure}[!t]
 \centering
 \includegraphics[angle=0,width=0.48\textwidth]{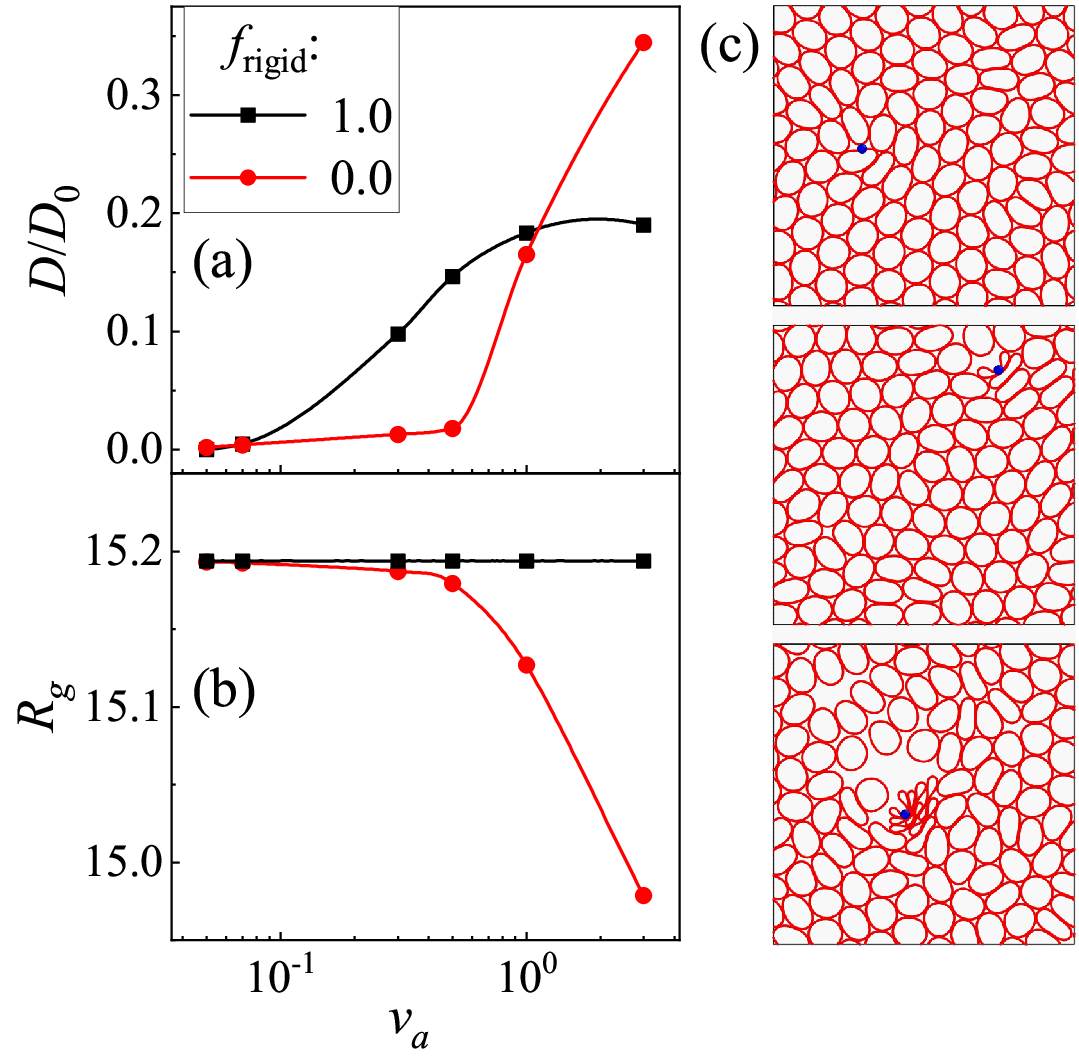}
 \caption{
 (a) Scaled diffusion coefficient of the active particle and (b) the radius of gyration of rings as functions of the active velocity $v_a$ for systems with purely rigid (black squares) and purely flexible (red circles) rings. (c) Snapshots of the active particle (blue disc with twice the diameter magnified) within purely flexible rings (red) at $v_a = 0.3, 1.0, 3.0$ (from top to bottom).
}
\label{fig:pure}
\end{figure}

To gain a deeper understanding of the non-monotonic dependence of active particle dynamics on the relative fraction of rigid versus flexible rings, we analyze the behavior of an active particle with $\tau_p=10^3$ in environments composed entirely of rigid ($f_{\rm rigid}=1.0$) or entirely flexible ($f_{\rm rigid}=0.0$) rings. We examine how the diffusion coefficient varies with active velocity for these two extreme cases, as shown in Fig.~\ref{fig:pure}(a). Here, $D$ is scaled by $D_0$, the diffusion coefficient of a single active particle in free space. In rigid rings, the dynamics of the active particle accelerates with increasing active velocity $v_a$, eventually reaching a plateau when $v_a$ becomes sufficiently large ($v_a>1.0$). The acceleration of the active particle indicates that it can squeeze through small gaps between neighboring rigid rings, likely due to its enhanced reorientation ability~\cite{JanusParticles,RN16,Lazaro2021,Luis2006}, which in turn improves the particle's efficiency in identifying potential channels within the rigid rings. At even higher $v_a$ values, the influence of the rigid rings becomes more dominant, resulting in the saturation of the $D/D_0$ at large $v_a$.

In flexible rings, however, the diffusion of the active particle exhibits a distinct dependence on $v_a$. At very low $v_a$ values ($\leq0.1$), the relative diffusion coefficient closely resembles that observed in rigid rings. This similarity explains the absence of non-monotonic dynamics in mixtures of rigid and flexible rings when varying the relative fraction of each component, as shown in Fig. S3 for $v_a=0.1$. Under these conditions, the active particle is unable to distinguish between rigid and flexible rings due to the low level of activity. As $v_a$ increases further, the diffusion coefficient initially remains nearly constant at a low value, followed by a sharp increase around $v_a=1.0$. This surge continues at higher $v_a$ values, where $D$ has already saturated in purely rigid rings. These observations suggest that the dynamics of active particles in flexible rings differ significantly from those in rigid rings. This difference may stem from the deformation of flexible rings, a property that can be influenced by the active particle.

To quantify the deformation of rings, we calculate the dependence of the gyration radius $R_g$ on $v_a$, as shown in Fig.~\ref{fig:pure}(b). For rigid rings, $R_g$ remains constant, as expected. In contrast, for flexible rings, $R_g$ matches the value of rigid rings at low $v_a$ but decreases rapidly at higher $v_a$. The $v_a$ at which $R_g$ decreases sharply closely corresponds to the point where the diffusion of the active particle increases significantly. This suggests that the acceleration in the dynamics of the active particle at high $v_a$ is due to conformational changes of flexible rings. To highlight the influence of the active particle on the deformation of flexible rings, we visualize the configurations at three selected values of $v_a$ ($0.3, 1.0$ and $3.0$). At large $v_a$ ($1.0$ and $3.0$), we observe significant deformation of the flexible rings near the active particle, with more rings deformed at $v_a=3.0$. The deformation of flexible rings creates vacancy spaces that facilitate the motion of the active particle when it reverses its moving direction.

\subsection*{Trapping processes}

\begin{figure}[!t]
 \centering
 \includegraphics[angle=0,width=0.48\textwidth]{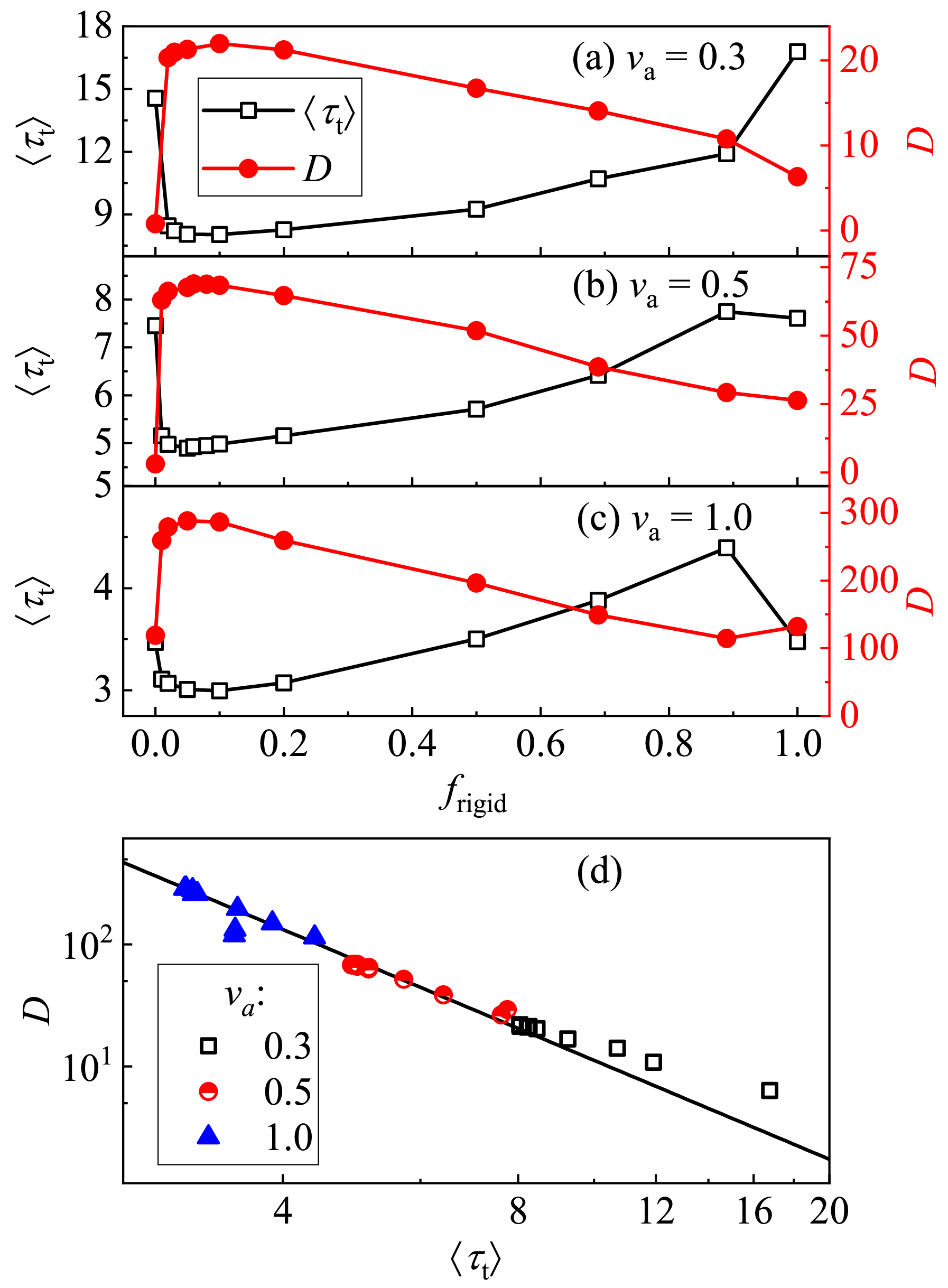}
 \caption{Average trapping duration $\langle\tau _{\rm{t}}\rangle$ (black, left axis) and diffusion coefficient $D$ (red, right axis) as functions of the fraction $f_{\rm rigid}$ of rigid rings for different particle active velocities (a) $v_a =0.3$,(b) 0.5, (c)1.0. (d) Diffusion coefficient $D$ as a function of $\langle\tau _{\rm{t}}\rangle$ at different values of ${v_a}$, as in the legend. The black solid line in (d) resents the power-law relationship $D\propto\langle\tau_{\rm t}\rangle^{-2.7}$.
\label{fig:transtime}
}
\end{figure}

The above observed different responses of active particle diffusion to rigid and flexible rings could lead to a non-monotonic dependence of diffusivity on the relative fraction of one component in a mixture of rigid and flexible rings. We directly examine the effect of the rings on hindering the active particle's motion by studying the particle's trapping duration, $\tau_{\rm t}$, defined as the duration until the active particle escapes the local cage formed by neighboring rings, a process referred to as a ``hop". The hop process is identified as occurring at the earliest time when at least one neighbor of the active particle, determined via the Voronoi method, deviates from its initial configuration reference. We present the average trapping duration $\langle \tau _{\rm{t}}\rangle$ (black open squares) along with the diffusion coefficient (red solid circles) as functions of $f_{\rm rigid}$ in Figs.~\ref{fig:transtime}(a)-\ref{fig:transtime}(c), for three different values of $v_a$ (0.3, 0.5 and 1.0) at fixed $\tau_p=10^3$. We find that $D$ and $\langle \tau _{\rm{t}}\rangle$ are closely related, with $\langle \tau_{\rm t}\rangle$ increasing as $f_{\rm rigid}$ increases when $D$ decreases, and vice versa. More intriguingly, $\langle \tau_{\rm t}\rangle$ exhibits a non-monotonic dependence on $f_{\rm rigid}$, mirroring the behavior of $D$ and displaying the same transition points. In Fig.~\ref{fig:transtime}(d), we further plot the diffusion coefficients $D$ as a function of $\langle\tau_{\rm t}\rangle$ for those three values of $v_a$. The data collapse well to a power-law function, $D\propto\langle\tau_{\rm t}\rangle^{-2.7}$. This suggests that the trapping by neighboring rings is the primary factor hindering the diffusive motion of the active particle.

\begin{figure}[!t]
 \centering
 \includegraphics[angle=0,width=0.48\textwidth]{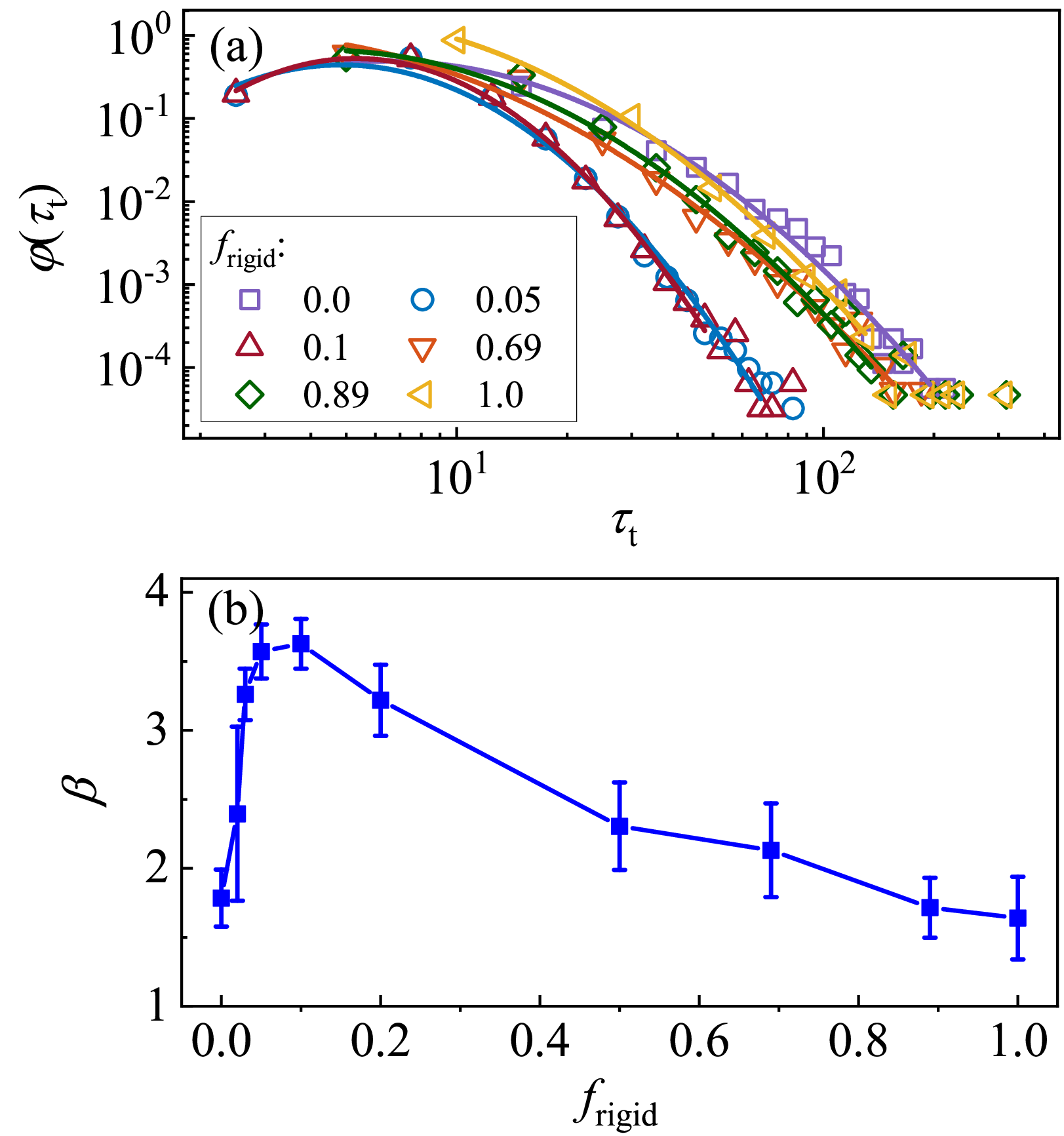}
 \caption{
(a) Probability distribution of the trapping duration $\tau_{\rm t}$. The open symbols show the numerical results while solid lines indicate the fitting from the extended entropic trap model (see text). (b) The fitting parameter $\beta$ for the system with $\tau_p=10^3$ and $v_a = 0.3$ as a function of $f_{\rm{rigid}}$. Here the error bar is from the standard deviations.
\label{fig:entropy}
}
\end{figure}

Previous studies on the trapping and hopping of bacteria in fixed porous media suggest that the probability distribution of trapping duration follows a power-law across the entire range of trapping durations~\cite{Jongyoon1999,Tapomoy2019,Christina2021}. However, our distribution, $\phi(\tau _{\rm{t}})$, presented in Fig.~\ref{fig:entropy}(a) (open symbols), deviates from this power-law behavior and instead exhibits a peak at small $\tau _{\rm{t}}$. To understand this phenomenon, we propose an updated version of the previously established entropic trap model, which has been used to explain the power-law behavior of the probability distribution of trapping duration~\cite{Jongyoon1999,Tapomoy2019,Christina2021}. In this model, the escape of active substances from the trapping in a fixed porous medium depends on the number of orientations that trap them, $\Omega_t$, and the number of orientations that allow their escape, $\Omega_e$. The difference in free energy between these two states defines the depth of the entropic trap, $C$, with its average value given by \cite{Christina2021} $C_0 \sim\langle\ln\frac{\Omega_t}{\Omega_e}\rangle$. Assume the probability density of trap depth $P(C)=C_0^{-1}\exp(-C/C_0)$ and considering the Arrhenius-like relationship between trapping duration and the depth of the entropic trap $\tau_t=\tau_0 \exp(C/X)$, where $\tau_0$ is a typical Brownian time scale and $X$ characterizes the ability of the particle to escape from the traps, one can derive a power-law distribution of $\tau_t$. Unlike the systems previously investigated~\cite{Jongyoon1999,Tapomoy2019,Christina2021}, the rings in our system are not fixed; they can move, and the flexible rings can even deform. The activity of particles, characterized by the parameter $X$, could influence the motion or deformation of the rings, leading to a reduction in the entropic depth $C$ by a fraction of $C/X$. Thus we assume a new distribution of the entropic depth $P(C)={\sqrt {\pi {C_0}X} }^{-1}\exp(-C^2/C_{0}X)$, which conforms to the Gaussian distribution. Consequently, the distribution of trapping time, $\varphi(\tau_t)$, is given by $P(C)(\partial\tau_t/\partial C)^{-1}$$=\frac{\beta}{\tau_0}\tau_t^{-1}e^{-C^2/C_0X}$$=\frac{\sqrt {\beta/\pi  }}{\tau_0}\frac{\tau_t}{\tau_0}^{-\beta\log\frac{\tau_t}{\tau_0}-1}$ with $\beta=X/C_0$ characterizing the competition between particle activity and confinement of rings. We plot the theoretical prediction of $\varphi(\tau_t)$ as lines in Fig.~\ref{fig:entropy}(a), showing good agreement with the numerical results (represented by symbols).

In Fig.~\ref{fig:entropy}(b), we present $\beta$ as a function of $f_{\rm rigid}$. Notably, $\beta > 1$ across the entire range of $f_{\rm rigid}$, which contrasts with results from fixed porous media, where $\beta$ can be less than 1~\cite{Christina2021}. This indicates that our medium are significantly influenced by active particles, which may create space and induce reorientation, leading to a reduction in the entropic depth. As a result, the active energy dominates over entropic effects. Moreover, $\beta$ exhibits a non-monotonic dependence on $f_{\rm rigid}$, mirroring the non-monotonic behavior observed in the diffusion coefficient and trapping time, as shown in Fig.~\ref{fig:msd}(b) and \ref{fig:transtime}(a). We find similar behavior at different values of $\tau_p$ and $v_a$, as illustrated by another example in supplementary Fig. S2 in the SI~\cite{SM}. This suggests that the reorientation enabling active particles to escape, or the influence of active effects on the migration of the particle, depend non-monotonically on the fraction of rigid versus flexible rings.

\subsection*{Theoretical prediction of optimal diffusion}
\begin{figure}[!t]
 \centering
 \includegraphics[angle=0,width=0.48\textwidth]{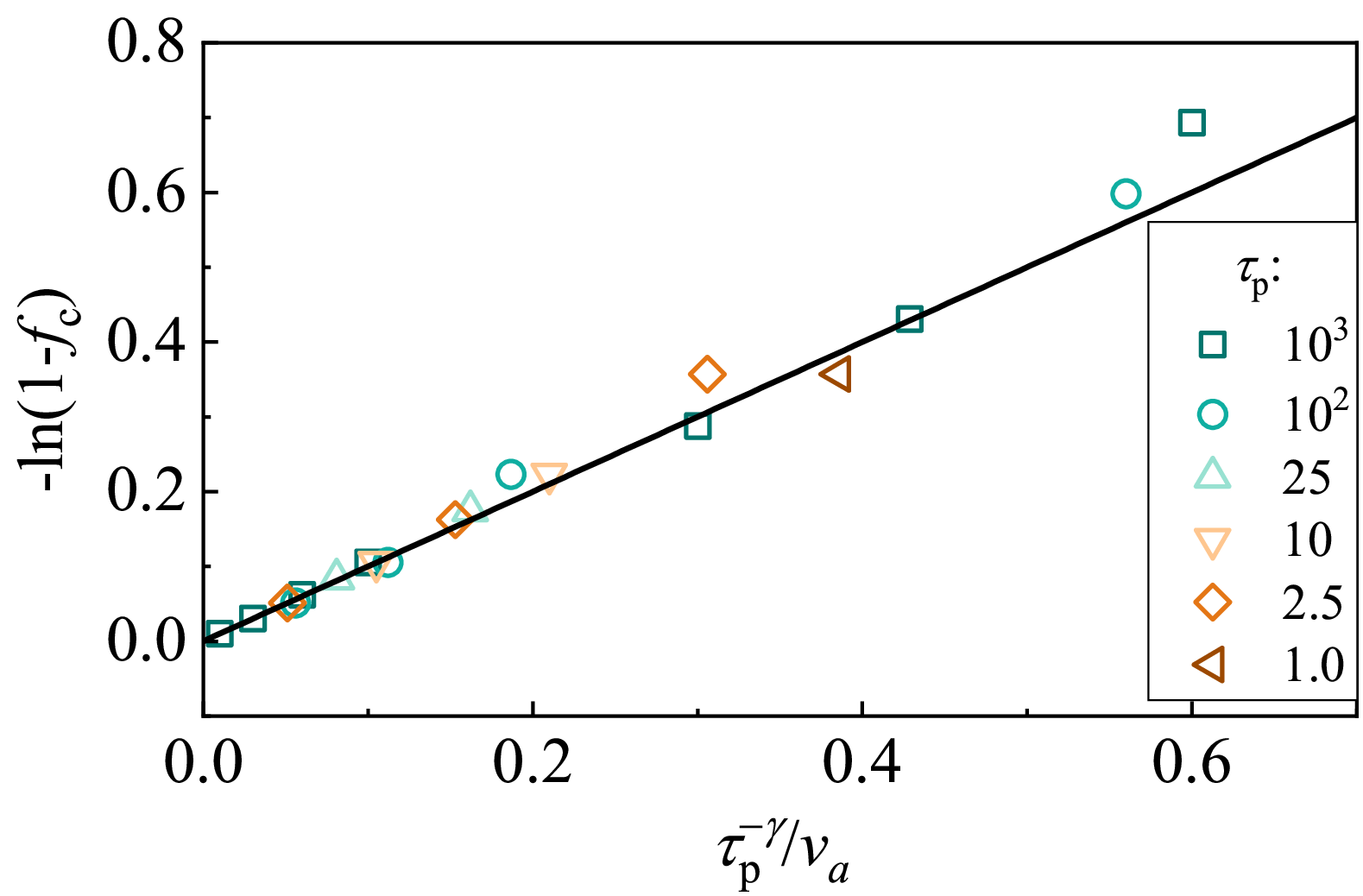}
 \caption{The optimal fraction $f_c$ of rigid rings for diffusion exhibits a universal relationship with the persistence time $\tau_p$ and the active velocity $v_a$. Symbols represent numerical results, while the black solid line denotes the theoretical prediction, given by $\ln (1-f_c) \propto \tau_p^{-\gamma} / v_a$.
\label{fig:scal}
}
\end{figure}

We finally theoretically explain the transition point in the observed non-monotonic dependence of the dynamics on $f_{\rm rigid}$. Flexible rings can deform, generating voids that facilitate the movement of active particles. We assume that each flexible ring contributes a unit length of space to the voids, while rigid rings contribute none. The size of the voids in fixed porous media has been shown to be crucial for the motion of active particles \cite{Christina2021}. In a disordered distribution of rigid and flexible rings, the probability that an active particle collides with flexible rings $n$ times is given by $(1-f_{\text {rigid}})^n$, and the corresponding length of the voids is expressed as $L(f_{\text {rigid}}, n)=n(1-f_{\text {rigid}})^n$. Given the long simulation time, we optimize $n$ to maximize $L_m(f_{\text {rigid}})=-\frac{1}{e \ln (1-f_{\text {rigid}})}$. We find that $L_m(f_c) \propto v_a \tau_p^\gamma$, leading to $\ln (1-f_c) \propto \tau_p^{-\gamma} / v_a$, where $\gamma \simeq 0.26$ and $f_{\rm c}$ represents the non-monotonic transition point. As shown in Fig.~\ref{fig:scal}, this relationship holds over a wide range of $\tau_p$ and $v_a$, offering a straightforward prediction of the non-monotonic transition point based on persistence and active velocity. The excellent agreement between the numerical results and theoretical predictions suggests that the non-monotonic dependence of the dynamics on $f_{\rm rigid}$ arises from the activity of the active particle. This activity influences the availability of voids, primarily generated by collisions with flexible rings, which allow the active particle to swim freely.
\section*{Conclusions}
In conclusion, an active particle in a slowly moving heterogeneous medium, consisting of a mixture of rigid and flexible rings, exhibits complex dynamics. Notably, the diffusivity shows a non-monotonic dependence on the relative proportion of rigid to flexible rings. This non-monotonic behavior arises when the particle's active velocity is sufficiently high, leading to the existence of an optimal diffusion of the particle within the medium. We speculate that rigid rings may enhance particle reorientation, improving its ability to explore potential hopping routes, while flexible rings modulate the local environment by creating void spaces that accelerate particle diffusion. In addition to diffusivity, we also observe a corresponding non-monotonic behavior in trapping duration, which follows a probability distribution predicted by the extended entropic trap model. The non-monotonic transition point, i.e., the optimal fraction $f_c$ of rigid to flexible rings for the spreading of an active particle, can be theoretically predicted by a universal relationship between $f_c$, the persistence time, and the active velocity: $\ln(1 - f_{\rm c}) \propto \tau_p^{-\gamma}/v_a$. This theoretical prediction aligns well with our simulation results and provides valuable insights into understanding and potentially application of the optimal spreading strategies for active particles in complex media. Our work could potentially be extended; for instance, tuning the density of the medium or the bond strength of flexible rings may shift the optimal diffusion point. These results may also inspire experimental studies on biological systems, such as the dynamics of bacteria in heterogeneous media.

\begin{acknowledgments}
Y.-W.L. acknowledges the support from the National Natural Science Foundation of China (NSFC) (Grant Nos. 12422501, 12374204 and 12105012). N.Z. acknowledges support from the NSFC (Grant No. 12475031).
\end{acknowledgments}


\end{document}